# White-light Quantitative Phase Imaging Unit

YoonSeok Baek,[1] KyeoReh Lee,[1] Jonghee Yoon,[1] Kyoohyun Kim[1], and YongKeun Park[1,2*]

[1]*Department of Physics, Korea Advanced Institutes of Science and Technology, Daejeon 34141, South Korea*
[2]*TomoCube, Inc., Daejeon 34051, Republic of Korea*
[*]*yk.park@kaist.ac.kr*

**Abstract:** We introduce the white light quantitative phase imaging unit (WQPIU) as a practical realization of quantitative phase imaging (QPI) on standard microscope platforms. The WQPIU is a compact stand-alone unit which measures sample induced phase delay under white-light illumination. It does not require any modification of the microscope or additional accessories for its use. The principle of the WQPIU based on lateral shearing interferometry and phase shifting interferometry provides a cost-effective and user-friendly use of QPI. The validity and capacity of the presented method are demonstrated by measuring quantitative phase images of polystyrene beads, human red blood cells, HeLa cells and mouse white blood cells. With speckle-free imaging capability due to the use of white-light illumination, the WQPIU is expected to expand the scope of QPI in biological sciences as a powerful but simple imaging tool.

**1. Introduction**

The optical transparency of biological samples has troubled biologists with low-contrast and indistinct edges. For this reason, phase-contrast imaging techniques, such as phase-contrast microscopy and differential interference contrast microscopy, have been widely adopted in biological sciences. Certainly phase-contrast imaging is sufficient in terms of visualization, yet it is limited to a qualitative description of morphology. Quantitative phase imaging (QPI) [1, 2] overcomes the limitation of phase-contrast imaging, providing quantitative information about biological samples including dry mass, membrane fluctuation, surface area, volume, and three-dimensional refractive index distribution [3-5]. Due to its label-free and quantitative imaging capability, QPI techniques have been utilized for the study of cell biology [6-12], tissue pathology [13, 14], and pathophysiology of diseases [15-18].

Despite these unique benefits, the transfer of QPI to biological and biomedical sciences is technically demanding. The difficulty is the limited compatibility of QPI with conventional optical microscopes due to bulky interferometric setups. Recently, there has been attempts to adopt QPI techniques of compact setups such as Michelson interferometry [19], quadriwave lateral shearing interferometry [20], quantitative phase imaging unit [21], $\tau$ interferometry [22, 23] and dual-channel interferometry [24]. But such techniques require coherent illumination which is an unfavorable requirement for standard bright-field microscope platforms. In this regard white light QPI techniques, such as white light diffraction phase microscopy [25], spatial light interference microscopy [26] and transport of intensity equation method [27-29], can be considered except that they require bulky and expensive setups, or microscope accessories with series of defocused images.

In this paper, we present a novel white light QPI technique highly compatible with standard bright-field microscopes. This technique, named as the white light quantitative phase imaging unit (WQPIU), requires neither bulky setup, additional equipment, nor modification of microscopes for its use. The WQPIU provides a compact interferometric design, employing the principle of lateral shearing interferometry [20, 30] and phase shifting interferometry [31]. The optical field is measured from four phase-shifted interferograms generated by a liquid crystal retarder. Furthermore, the use of existing white-light illumination also solves speckle problems in conventional QPI techniques with coherent lasers. We demonstrate the capacity of the present method by measuring quantitative phase images of polystyrene beads, human red blood cells (RBCs), human uterine cervical carcinoma cells (HeLa cells), and mouse macrophages.

**2. Materials and Methods**

*2.1. Setup and Principle*

The schematic of the WQPIU placed at an output port of a microscope is illustrated in Fig. 1. The WQPIU is composdddddaaed of a beam displacer, a liquid crystal retarder, and an optical path difference (OPD) compensator and a linear polarizer. The beam displacer (United Crystals, New York, USA) is a birefringent material whose optic axis lies at 45° with respect to its surface normal. The optic axis is shown in the black double-headed arrow in Fig. 1. The beam displacer separates light into two orthogonally polarized lights: one parallel to the optic axis (e-ray) and the other perpendicular to the optic axis (o-ray). The wavefronts of resultant e-ray and o-ray are replicas of each other. The propagation direction of light remains the same before and after the beam displacer. Therefore, e-ray and o-ray propagate in parallel to each other and to the incident light. As a result, the image of o-ray (or e-ray) beam overlaps the laterally translated image of e-ray (or o-ray). When a specimen contains sparse samples so that the o-ray and e-ray can be used as a sample and a reference beam in lateral shearing interferometry, the sample induced phase can be obtained by measuring the relative phase between two beams. In addition the liquid crystal retarder (LCC1223-A, Thorlabs., New Jersey, USA) was adopted. The liquid crystal retarder modulates the phase difference between the o-ray and e-ray, which enables phase shifting interferometry.

However, the interference between the two beams is assured when the OPD is comparable to the temporal coherence length. Since e-ray and o-ray undergo different optical path in the beam displacer, this OPD must be compensated to assure the interference between two beams. In particular, when temporally incoherent light is used such as in this case, the minimization of the OPD is crucial in order to achieve interference. For this reason, the OPD compensator was devised. The OPD compensator (United Crystals, New York, USA) is a birefringent material whose optics axis is perpendicular to its surface normal. It gives a polarization dependent phase delays whose difference cancels the OPD at the beam displacer. In addition, for two laterally separated beams of the identical wavefront to interfere, the light must have a spatial coherence length greater than the beam separation at the beam displacer. Such spatially coherent light can be achieved with a bright-field microscope either by reducing the size of illumination aperture or by placing the light source far from the sample.

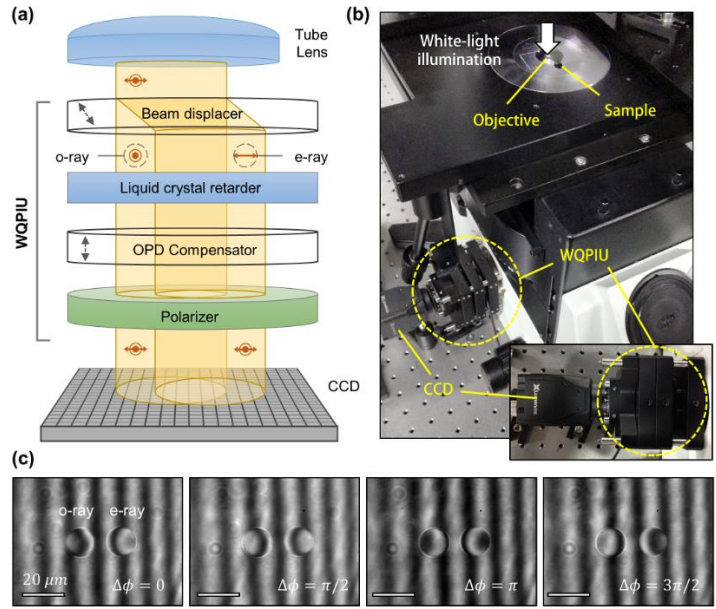

Fig. 1. (a) The schematic of the WQPIU. The WQPIU is composed of a beam displacer, a liquid crystal variable retarder, an OPD compensator and a linear polarizer. Black arrows indicate optic axis of each crystal. Orange arrows and dots indicate the polarization of light (b) A photograph of the WQPIU attached to a microscope. (Inset) A photograph of the WQPIU and the CCD (c) Phase shifted interferograms recorded by the CCD. Total 4 images are recorded with phase delays of 0, π/2, π, 3π/2 respectively.

After the linear polarizer (LPVISE100-A, Thorlabs, New Jersey, USA), two images of a sample with interference pattern are observed (Fig. 1(c)). The vertical lines indicate that the wavefronts of two beams are not ideally flat but tilted; however, this tilt can be easily removed because it is independent of a sample. Then the optical field at the sample plane can be retrieved using phase shifted interferograms as [32],

$$E(x, y)e^{i\phi(x,y)} = \frac{I_0(x, y) - I_\pi(x, y) - i[I_{\pi/2}(x, y) - I_{3\pi/2}(x, y)]}{4} \quad (1)$$

where subscripts indicate relative phase delay between sample beam and reference beam. Taking the complex argument, a quantitative phase image can be obtained.

Artifacts such as stains and dusts on the image were eliminated by subtracting the background phase images from the sample phase images. The whole process of data acquisition was automated by custom LabVIEW software. A typical run time of the software was less than 700 ms, where the speed limiting step is the control and response of the liquid crystal.

## 2.2. Beam separation and OPD compensation

The separation of the beam displacer must be larger than the size of the sample image to avoid overlap between sample images. The separation can be calculated using material refractive indices and thickness. For the sake of compactness, we chose calcite which has strong birefringence as a material. The designed beam displacer has a separation of 0.9 mm, which corresponds to 15 μm at the sample plane under 60× magnification, at 550 nm. The thickness of such beam displacer is 8.15 mm. The corresponding separation of the beam displacer over the visible spectrum is shown in Fig. 2(a).

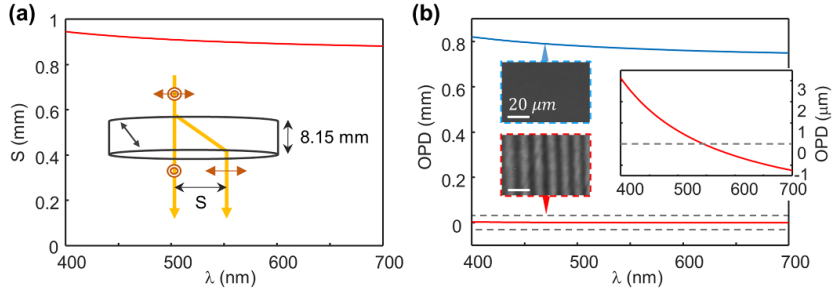

Fig. 2. Separation (S) at the beam displacer and OPD. (a) The beam separation at the Beam displacer. The separation varies depending on the wavelength due to the dispersion of calcite. (b) OPD between o-ray and e-ray at normal incidence. The upper blue curve shows the OPD at the beam displacer. The lower red curve shows OPD after both the beam displacer and the compensator. (Inset) CCD images with corresponding OPDs and a magnified graph of the lower red curve. OPD is reduced to few micrometers at the visible range after the compensation.

The separation varies depending on the wavelength due to the dispersion of calcite. The difference in separations between 400 nm and 700 nm is about 0.06 mm, which results in lateral chromatic aberration. However this only applies to e-ray, since o-ray does not undergo lateral displacements at normal incidence. Thus to minimize the lateral chromatic aberration, the sample images of o-ray were used throughout the experiments.

The calculated OPD at the beam displacer is shown in Fig. 2(b) as a blue curve. The OPD at the beam displacer is about 0.8 mm over the visible spectrum. No sign of interference between two beams was found right after the beam displacer. To minimize the OPD, the OPD compensator was designed. The thickness of the compensator was determined to make OPD at 550 nm be zero. The used OPD compensator is made of calcite having a thickness of 4.41 mm. After the compensation, the OPD is reduced by a factor of 1 000 as shown in Fig. 2(b), and the interference between two beams is clearly observed.

## 2.3. Liquid Crystal Calibration

Since the liquid crystal retarder does not give uniform phase delay at every wavelength, it must be calibrated beforehand. The calibration was performed with two linear polarizers. First the liquid crystal retarder was placed between two linear polarizers, and then the transmitted intensity was measured by changing the voltage applied to the liquid crystal. From this, the retardation at each voltage is obtained.

## 3. Results and Discussion

In the experiments, an inverted microscope (IX71, Olympus Inc., Japan) and a xenon lamp (U-LH75XEAPO, Olympus Inc., Japan) were used. A condenser was removed to secure spatial coherence at the sample plane. The WQPIU and a monochromatic camera (Lt365R, Lumenera Inc., Canada) were placed at an output port of the microscope as depicted in Fig. 1(b).

To demonstrate the validity of the WQPIU, we first measured the phase image of 10 μm polystyrene beads for verification. The polystyrene beads ($n$ = 1.5959 at $\lambda$ = 550 nm) was immersed in index matching oil ($n$ = 1.5532 at $\lambda$ = 550 nm) and was sandwiched by two cover glasses (Marienfeld-Superior Inc., Germany). A 60× objective lens (PLAPON 60XO, Olympus Inc., Japan) was used for imaging. The measured quantitative phase image of the bead is shown in Fig. 3(a). The spherical shape of the bead is fully depicted in the phase image and the line profile along the dashed line, verifying the working principle of the WQPIU.

Next, to demonstrate the capacity and applicability of the WQPU in bioimaging, we measured the phase of human RBCs. The RBCs were diluted 200 times with Dulbecco's Phosphate Buffered Saline (DPBS, Welgene Inc., Republic of Korea) and sandwiched by two cover glasses according to the sample preparation protocol [17]. The sample preparation protocol was approved by the Institutional Review Board (IBR project number: KH2015−37). Again the same 60× objective lens (PLAPON 60XO, Olympus Inc., Japan) was used for imaging. The dimple shape of RBC is observed in the result shown in Fig. 3(b).

Further demonstration of the biological application of the WQPIU was conducted on HeLa cells. HeLa cells were maintained in Dulbecco's Modified Eagle's Medium (DMEM, Welgene Inc., Republic of Korea) supplemented with 10% heat-inactivated fetal bovine serum, 1 000 U/L penicillin, and 1 000 μg/mL streptomycin in a humidified incubator at 37°C and 5% $CO_2$. Cells were plated on a 24×40 mm cover glass (Marienfeld-Superior Inc., Germany) and maintained for 4−8 hours in an incubator. Then cells were washed with DPBS solution immediately before the experiments. Next, 200 μL of prewarmed DMEM was added to the sample, and then another cover glass was placed on the sample to prevent the sample drying. The prepared HeLa cells were imaged by 20× objective (LMPLFLN 20X, Olympus Inc., Japan). The obtained phase image is shown in Fig. 3(c) where the morphology of the cell along with its internal structures is well observed.

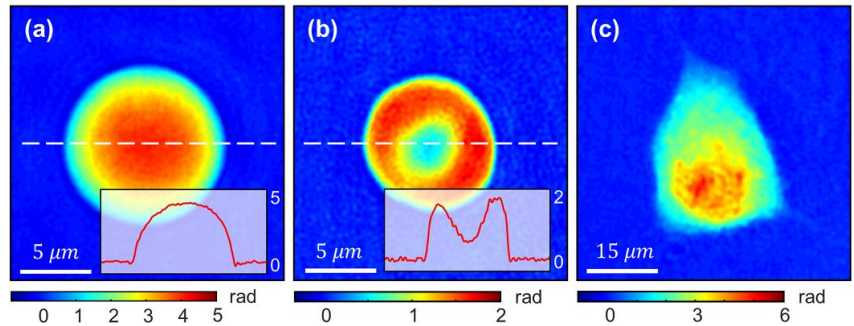

fig. 3. Quantitative phase images of various samples. (a) 10 μm Polystyrene bead immersed in index matching oil ($n$ = 1.5532 at $\lambda$ = 550 nm) and its line profile along the white dash line (b) RBC and its line profile along the white dash line (c) HeLa cell

Next to demonstrate the suppression of speckle noise, we compared quantitative phase images of 10 μm polystyrene beads under coherent illumination and under incoherent illumination. Two light sources, a 532 nm DPSS laser (SambaTM, Cobolt Inc, Sweden) and a xenon lamp (U-LH75XEAPO, Olympus Inc., Japan), were used. Polystyrene beads were immersed in index matching oil ($n$ = 1.5633 at $\lambda$ = 550 nm) and sandwiched by two cover glasses. The corresponding results are shown in Fig. 4(a) and Fig. 4(b). Incoherent illumination shows significantly suppressed coherent noise with a clear sample image. In contrast, coherent illumination shows considerable ripple-like coherent noise distorting the sample phase. The phase distribution at the lower-right corners of the images are shown in the histogram (Fig. 4(c)). The spatial standard deviation of phase images of incoherent illumination ($\sigma$ = 0.0273 rad) is smaller by a factor of 9 compared to that of coherent illumination ($\sigma$ = 0.2462 rad).

In case of coherent illumination, the major contribution to the noise comes from multiple reflection between cover glasses. In practical cases where mechanical vibration is significant, such coherent noise will increase inevitably. However, the WQPIU is free of such coherent noise and its phase sensitivity is determined by shot noise and dynamic rages of the camera and the liquid crystal retarder.

Finally to demonstrate the capacity of the WQPIU in live-cell dynamic imaging, we measured quantitative phase images of mouse macrophages at a frame rate of 1.47 Hz (Visualization 1 & 2). Macrophages were collected from mice peritoneal cavity according to a sample preparation protocol [33]. In brief, 7-weeks-old Balb/c mouse (Orient Bio Inc., Republic of Korea) was euthanized with $CO_2$ gas. Then 5 mL of ice–cold DPBS (with 3% FBS) was injected into the peritoneal cavity of euthanized mice using a syringe with a 25G needle. Injected fluid was collected from peritoneal cavity using a syringe with a 26G needle and centrifuged at 250 g for 8 min. The macrophages resuspended in DMEM and maintained in a humidified incubator at 37°C and 5% $CO_2$ for 2 days before experiments. The sample preparation protocol was approved by the Institutional Review Board (IBR project number: KA−2015−03).

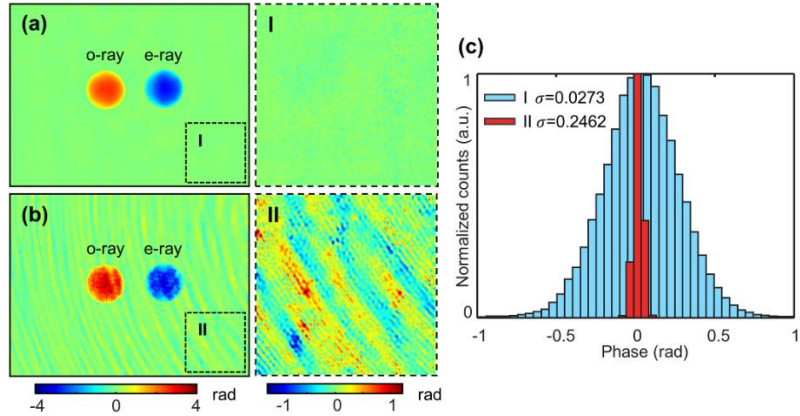

Fig. 4. Quantitative phase images under different illuminations. The same 10 μm polystyrene bead immersed in index matching oil (n = 1.5633 at $\lambda$ = 550 nm) was imaged (a) under xenon lamp illumination and (b) under 532 nm DPSS laser. (c) A phase histogram of two background regions denoted as I and II.

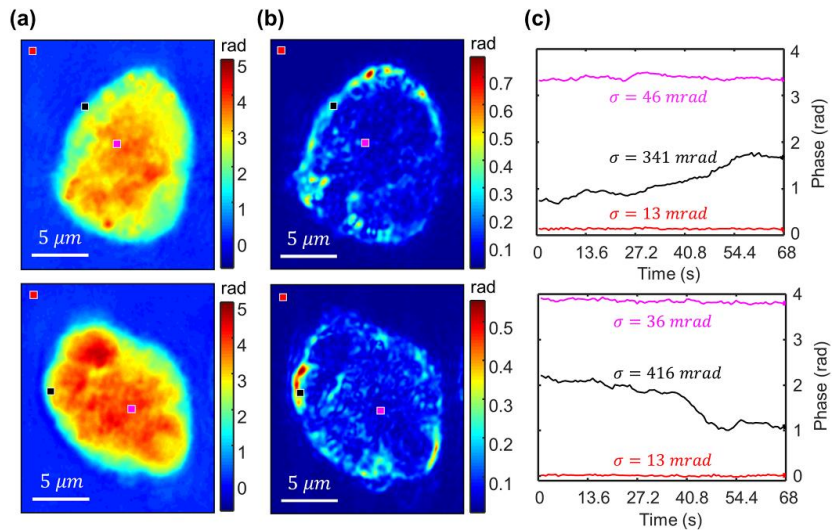

Fig. 5. Two mouse macrophages were observed at a frame rate of 1.47 Hz. (a) The phase images of the samples at t=0. (b) Phase standard deviations over 100 frames (c) Phase values at the different points marked with red, black and magenta boxes.

The measured phases of each sample at *t* = 0 are shown in Fig. 5(a) and temporal standard deviations over 100 frames are shown in Fig. 5(b). We examined the changes in phase valued at different points over time. Three points at background, boundary of the cell and inside the cell were investigated. The phase changes at the boundaries are gradual, which indicates the slow movements of the cells toward or away from the point. The phases inside the cells are relatively stable and yet even the minute changes are detected. The background point shows a flat distribution with standard deviation of 13 mrad. In general, the background standard deviations are 14 ± 2 mrad. This measured background stability is comparable with, yet slightly higher than, that of compact QPI techniques using coherent light sources [21, 24, 34]. The noise from mechanical vibration is minimized by the common-path setup and high-frequency noise faster than the liquid crystal retarder response is considered relevant.

The data acquisition took 170 ms for each interferogram and 680 ms for one quantitative phase image throughout the experiments. The faster acquisition can also be achieved by optimizing LabVIEW code, by adopting faster equipment or by implementing other phase-shifting algorithms such as two-step or three-step methods [35, 36]. Although the principle of the WQPIU assumes plane incident wavefront, the WQPIU operates successfully even at quasi-plane wavefronts. The wavefront magnified by an optical microscope is enough to be used, hence no spatial filtering of illumination is required.

## 4. Conclusion

We developed the WQPIU as a cost-effective and non-expert friendly white light QPI technique. The WQPIU is composed of four optical components that are easily available and can be put into a compact unit. In addition, the WQPIU does not require hardware modification or microscope accessories to be employed. Thus, it can utilize the ubiquitous-ness of conventional bright-field microscopes. Its principle based on lateral-shearing interferometry with phase-shifting interferometry has been validated through the experiments. The high spatio-temporal stability of the WQPIU has also been confirmed. The demonstration on biological cells indicates the potential of the WQPIU in biological applications including live-cell imaging. Its phase sensitivity is limited by hardware noise and faster data acquisition can be accomplished by efficient programing and different phase-retrieval methods. Although the WQPIU requires sparse samples, we believe this condition is minor when dilution of the sample is possible, or the condition can be relaxed by changing the magnification and using a beam displacer with larger beam separation. The WQPIU allows practical use of various analysis including white light 3-D optical tomography [37], Fourier transform light scattering (FTLS) [38], spectroscopic [39, 40] and birefringent phase measurements [41]. We are positive that the WQPIU can further the access to QPI techniques and their potential applications in biological and medical fields.

## Acknowledgments

This work was supported by KAIST, and the National Research Foundation of Korea (2015R1A3A2066550, 2014K1A3A1A09063027, 2012-M3C1A1-048860, 2014M3C1A3052537) and Innopolis foundation (A2015DD126).